\documentclass[twocolumn]{aa}

\usepackage{graphicx}
\usepackage{amsmath,amsfonts,amssymb}
\usepackage{txfonts}
\usepackage{color}
\usepackage{natbib}
\usepackage{float}
\usepackage{dblfloatfix}
\usepackage{afterpage}
\usepackage{ifthen}
\usepackage[morefloats=12]{morefloats}
\usepackage{placeins}
\usepackage{multicol}
\bibpunct{(}{)}{;}{a}{}{,}
\usepackage[switch]{lineno}
\definecolor{linkcolor}{rgb}{0.6,0,0}
\definecolor{citecolor}{rgb}{0,0,0.75}
\definecolor{urlcolor}{rgb}{0.12,0.46,0.7}
\usepackage[breaklinks, colorlinks, urlcolor=urlcolor,
    linkcolor=linkcolor,citecolor=citecolor,pdfencoding=auto]{hyperref}
\hypersetup{linktocpage}
\usepackage{bold-extra}
\usepackage[export]{adjustbox}

\usepackage{cuted}

\def\setsymbol#1#2{\expandafter\def\csname #1\endcsname{#2}}
\def\getsymbol#1{\csname #1\endcsname}

\def\Planck{\textit{Planck}}

\newbox\tablebox    \newdimen\tablewidth
\def\leaderfil{\leaders\hbox to 5pt{\hss.\hss}\hfil}
\def\tablenote#1 #2\par{\begingroup \parindent=0.8em
    \abovedisplayshortskip=0pt\belowdisplayshortskip=0pt
    \noindent
    $$\hss\vbox{\hsize\tablewidth \hangindent=\parindent \hangafter=1 \noindent
    \hbox to \parindent{$^#1$\hss}\strut#2\strut\par}\hss$$
    \endgroup}

\def\L2{\ifmmode L_2\else $L_2$\fi}

\def\DeltaT{\ifmmode \Delta T\else $\Delta T$\fi}
\def\deltat{\ifmmode \Delta t\else $\Delta t$\fi}
\def\fknee{\ifmmode f_{\rm knee}\else $f_{\rm knee}$\fi}
\def\Fmax{\ifmmode F_{\rm max}\else $F_{\rm max}$\fi}
\def\solar{\ifmmode{\rm M}_{\mathord\odot}\else${\rm M}_{\mathord\odot}$\fi}
\def\Msolar{\ifmmode{\rm M}_{\mathord\odot}\else${\rm M}_{\mathord\odot}$\fi}
\def\Lsolar{\ifmmode{\rm L}_{\mathord\odot}\else${\rm L}_{\mathord\odot}$\fi}
\def\inv{\ifmmode^{-1}\else$^{-1}$\fi}
\def\mo{\ifmmode^{-1}\else$^{-1}$\fi}
\def\sup#1{\ifmmode ^{\rm #1}\else $^{\rm #1}$\fi}
\def\expo#1{\ifmmode \times 10^{#1}\else $\times 10^{#1}$\fi}
\def\,{\thinspace}
\def\lsim{\mathrel{\raise .4ex\hbox{\rlap{$<$}\lower 1.2ex\hbox{$\sim$}}}}
\def\gsim{\mathrel{\raise .4ex\hbox{\rlap{$>$}\lower 1.2ex\hbox{$\sim$}}}}

\def\simprop{\mathrel{\raise .4ex\hbox{\rlap{$\propto$}\lower 1.2ex\hbox{$\sim$}}}}
\def\deg{\ifmmode^\circ\else$^\circ$\fi}
\def\pdeg{\ifmmode $\setbox0=\hbox{$^{\circ}$}\rlap{\hskip.11\wd0 .}$^{\circ}
          \else \setbox0=\hbox{$^{\circ}$}\rlap{\hskip.11\wd0 .}$^{\circ}$\fi}
\def\arcs{\ifmmode {^{\scriptstyle\prime\prime}}
          \else $^{\scriptstyle\prime\prime}$\fi}
\def\arcm{\ifmmode {^{\scriptstyle\prime}}
          \else $^{\scriptstyle\prime}$\fi}
\newdimen\sa  \newdimen\sb
\def\parcs{\sa=.07em \sb=.03em
     \ifmmode \hbox{\rlap{.}}^{\scriptstyle\prime\kern -\sb\prime}\hbox{\kern -\sa}
     \else \rlap{.}$^{\scriptstyle\prime\kern -\sb\prime}$\kern -\sa\fi}
\def\parcm{\sa=.08em \sb=.03em
     \ifmmode \hbox{\rlap{.}\kern\sa}^{\scriptstyle\prime}\hbox{\kern-\sb}
     \else \rlap{.}\kern\sa$^{\scriptstyle\prime}$\kern-\sb\fi}
\def\ra[#1 #2 #3.#4]{#1\sup{h}#2\sup{m}#3\sup{s}\llap.#4}
\def\dec[#1 #2 #3.#4]{#1\deg#2\arcm#3\arcs\llap.#4}
\def\deco[#1 #2 #3]{#1\deg#2\arcm#3\arcs}
\def\rra[#1 #2]{#1\sup{h}#2\sup{m}}

\def\dots{\relax\ifmmode \ldots\else $\ldots$\fi}
\def\WHzsr{\ifmmode $W\,Hz\mo\,sr\mo$\else W\,Hz\mo\,sr\mo\fi}
\def\mHz{\ifmmode $\,mHz$\else \,mHz\fi}
\def\GHz{\ifmmode $\,GHz$\else \,GHz\fi}
\def\mKs{\ifmmode $\,mK\,s$^{1/2}\else \,mK\,s$^{1/2}$\fi}
\def\muKs{\ifmmode \,\mu$K\,s$^{1/2}\else \,$\mu$K\,s$^{1/2}$\fi}
\def\muKRJs{\ifmmode \,\mu$K$_{\rm RJ}$\,s$^{1/2}\else \,$\mu$K$_{\rm RJ}$\,s$^{1/2}$\fi}
\def\muKHz{\ifmmode \,\mu$K\,Hz$^{-1/2}\else \,$\mu$K\,Hz$^{-1/2}$\fi}
\def\MJysr{\ifmmode \,$MJy\,sr\mo$\else \,MJy\,sr\mo\fi}
\def\MJysrmK{\ifmmode \,$MJy\,sr\mo$\,mK$_{\rm CMB}\mo\else \,MJy\,sr\mo\,mK$_{\rm CMB}\mo$\fi}
\def\microns{\ifmmode \,\mu$m$\else \,$\mu$m\fi}

\def\muK{\ifmmode \,\mu$K$\else \,$\mu$\hbox{K}\fi}
\def\microK{\ifmmode \,\mu$K$\else \,$\mu$\hbox{K}\fi}
\def\muW{\ifmmode \,\mu$W$\else \,$\mu$\hbox{W}\fi}
\def\kms{\ifmmode $\,km\,s$^{-1}\else \,km\,s$^{-1}$\fi}
\def\kmsMpc{\ifmmode $\,\kms\,Mpc\mo$\else \,\kms\,Mpc\mo\fi}
\providecommand{\sorthelp}[1]{}

\def\WMAP{\textit{WMAP}}

\def\nside{N_{\mathrm{side}}}

\def\healpix{\texttt{HEALPix}}
\def\commander{\texttt{Commander}}

\renewcommand{\d}[0]{\vec{d}}

\newcommand{\n}[0]{\vec{n}}

\newcommand{\s}[0]{\vec{s}}
\renewcommand{\a}[0]{\vec{a}}

\newcommand{\T}[0]{\tens{T}}

\renewcommand{\L}[0]{\tens{L}}

\newcommand{\N}[0]{\tens{N}}

\renewcommand{\r}[0]{\vec{r}}

\newcommand{\BP}{\textsc{BeyondPlanck}}
\newcommand{\lfi}[0]{LFI}

\def\inv{^{-1}}

\begin{document}

\title{\bfseries{\scshape{BeyondPlanck}} XVI. Limits on large-scale polarized anomalous \\microwave emission from \Planck\ LFI and \WMAP}
\newcommand{\oslo}[0]{1}
\newcommand{\princeton}[0]{2}
\newcommand{\milanoA}[0]{3}
\newcommand{\milanoB}[0]{4}
\newcommand{\triesteB}[0]{5}
\newcommand{\planetek}[0]{6}
\newcommand{\jpl}[0]{7}
\newcommand{\helsinkiA}[0]{8}
\newcommand{\helsinkiB}[0]{9}
\newcommand{\nersc}[0]{10}
\newcommand{\milanoC}[0]{11}
\newcommand{\haverford}[0]{12}
\newcommand{\mpa}[0]{13}
\newcommand{\triesteA}[0]{14}
\author{\small
\textcolor{black}{D.~Herman}\inst{\oslo}\thanks{Corresponding author: D. Herman; \url{d.c.herman@astro.uio.no}}
\and
\textcolor{black}{B.~Hensley}\inst{\princeton}
\and
K.~J.~Andersen\inst{\oslo}
\and
\textcolor{black}{R.~Aurlien}\inst{\oslo}
\and
\textcolor{black}{R.~Banerji}\inst{\oslo}
\and
M.~Bersanelli\inst{\milanoA, \milanoB}
\and
S.~Bertocco\inst{\triesteB}
\and
M.~Brilenkov\inst{\oslo}
\and
M.~Carbone\inst{\planetek}
\and
L.~P.~L.~Colombo\inst{\milanoA}
\and
H.~K.~Eriksen\inst{\oslo}
\and
\textcolor{black}{M.~K.~Foss}\inst{\oslo}
\and
\textcolor{black}{U.~Fuskeland}\inst{\oslo}
\and
S.~Galeotta\inst{\triesteB}
\and
M.~Galloway\inst{\oslo}
\and
S.~Gerakakis\inst{\planetek}
\and
E.~Gjerl{\o}w\inst{\oslo}
\and
M.~Iacobellis\inst{\planetek}
\and
M.~Ieronymaki\inst{\planetek}
\and
\textcolor{black}{H.~T.~Ihle}\inst{\oslo}
\and
J.~B.~Jewell\inst{\jpl}
\and
\textcolor{black}{A.~Karakci}\inst{\oslo}
\and
E.~Keih\"{a}nen\inst{\helsinkiA, \helsinkiB}
\and
R.~Keskitalo\inst{\nersc}
\and
G.~Maggio\inst{\triesteB}
\and
D.~Maino\inst{\milanoA, \milanoB, \milanoC}
\and
M.~Maris\inst{\triesteB}
\and
S.~Paradiso\inst{\milanoA}
\and
B.~Partridge\inst{\haverford}
\and
M.~Reinecke\inst{\mpa}
\and
A.-S.~Suur-Uski\inst{\helsinkiA, \helsinkiB}
\and
T.~L.~Svalheim\inst{\oslo}
\and
D.~Tavagnacco\inst{\triesteB, \triesteA}
\and
H.~Thommesen\inst{\oslo}
\and
I.~K.~Wehus\inst{\oslo}
\and
A.~Zacchei\inst{\triesteB}
}
\institute{\small
Institute of Theoretical Astrophysics, University of Oslo, Blindern, Oslo, Norway\goodbreak
\and
Department of Astrophysical Sciences, Princeton University, Princeton, NJ 08544,
U.S.A.\goodbreak
\and
Dipartimento di Fisica, Universit\`{a} degli Studi di Milano, Via Celoria, 16, Milano, Italy\goodbreak
\and
INAF/IASF Milano, Via E. Bassini 15, Milano, Italy\goodbreak
\and
INAF - Osservatorio Astronomico di Trieste, Via G.B. Tiepolo 11, Trieste, Italy\goodbreak
\and
Planetek Hellas, Leoforos Kifisias 44, Marousi 151 25, Greece\goodbreak
\and
Jet Propulsion Laboratory, California Institute of Technology, 4800 Oak Grove Drive, Pasadena, California, U.S.A.\goodbreak
\and
Department of Physics, Gustaf H\"{a}llstr\"{o}min katu 2, University of Helsinki, Helsinki, Finland\goodbreak
\and
Helsinki Institute of Physics, Gustaf H\"{a}llstr\"{o}min katu 2, University of Helsinki, Helsinki, Finland\goodbreak
\and
Computational Cosmology Center, Lawrence Berkeley National Laboratory, Berkeley, California, U.S.A.\goodbreak
\and
INFN, Sezione di Milano, Via Celoria 16, Milano, Italy\goodbreak
\and
Haverford College Astronomy Department, 370 Lancaster Avenue,
Haverford, Pennsylvania, U.S.A.\goodbreak
\and
Max-Planck-Institut f\"{u}r Astrophysik, Karl-Schwarzschild-Str. 1, 85741 Garching, Germany\goodbreak
\and
Dipartimento di Fisica, Universit\`{a} degli Studi di Trieste, via A. Valerio 2, Trieste, Italy\goodbreak
}

\authorrunning{Daniel Herman}
\titlerunning{Constraints on polarized AME}

\abstract{We constrained the level of polarized anomalous microwave
  emission (AME) on large angular scales using \Planck\ Low-Frequency Instrument (LFI) and
  \WMAP\ polarization data within a Bayesian cosmic microwave background (CMB) analysis
  framework. We modeled synchrotron emission with a power-law spectral
  energy distribution, as well as the sum of AME and thermal dust emission
  through linear regression with the \textit{Planck} High-Frequency Instrument (HFI) 353 GHz
  data. This template-based dust emission model allowed us to constrain
  the level of polarized AME while making minimal assumptions on its
  frequency dependence. We neglected cosmic microwave background fluctuations,
  but show through
  simulations that these fluctuations have a minor impact on the results. We
  find that the resulting AME polarization fraction confidence limit
  is sensitive to the polarized synchrotron spectral index prior. In addition,
  for prior means $\beta_{\mathrm{s}} <$ $-3.1$ we find an
  upper limit of $p_{\mathrm{AME}}^{\rm max}\lesssim 0.6\,\%$ (95\,\%
  confidence). In contrast, for means $\beta_{\mathrm{s}}=-3.0$, we
  find a nominal detection of $p_{\mathrm{AME}}=2.5\pm1.0\,\%$ (95\,\%
  confidence). These data are thus not strong enough to simultaneously
  and robustly constrain both polarized synchrotron emission and AME,
  and our main result is therefore a constraint on the AME
  polarization fraction explicitly as a function of
  $\beta_\mathrm{s}$. Combining the current \textit{Planck} and
  \textit{WMAP} observations with measurements from high-sensitivity
  low-frequency experiments such as C-BASS and QUIJOTE will be
  critical to improve these limits further.  }


\maketitle

\section{Introduction}
\label{sec:introduction}

One of the primary goals of future cosmic microwave background (CMB) missions such as CMB-S4 \citep{cmbs4}, Simons Observatory \citep{SO2019}, and \textit{LiteBIRD} \citep{litebird2020} is to detect the hypothesized signal from primordial gravitational waves, or to constrain the tensor-to-scalar ratio $r$. In order for this goal to be reached, it is essential to characterize polarized Galactic emission at the 10--100~${\rm nK}$ level.

Two emission mechanisms dominate the polarized Galactic signal at microwave frequencies: synchrotron emission below $\sim$70\,GHz and thermal dust emission above $\sim$70\,GHz \citep[e.g.,][]{page2007,planck2014-a12}. Recent measurements from \Planck\ have shown that the polarization fractions of these two components can reach 40 and 20\,\%, respectively \citep{Planck2018xii}, and their rich morphologies are connected by the Galactic magnetic field. Other emission mechanisms, such as free-free and anomalous microwave emission (AME), are important contributions to the total intensity observed in the same frequency range, but no significant polarization has been detected yet \citep[e.g.,][]{planck2014-a31}, although a small free-free contribution is predicted on very small angular scales near the edges of \ion{H}{ii} regions \citep{rybicki_lightman}.

Anomalous microwave emission was serendipitously detected while analyzing \emph{COBE}--DMR and other data sets in the 10--60\,GHz range \citep{kogut1996,leitch:1997,deOliveira-Costa2004}. This emission is highly correlated with Galactic thermal dust morphology and has been detected within specific clouds such as the Perseus molecular cloud \citep{bob2005,planck_low_freq_2015,vsa_ame,QUIJOTE_I}, $\lambda$ Orionis \citep{lambdaO_AME,lambdaO_cbass_quijote}, $\rho$ Ophiuchus \citep{casassus2008,arce-tord2020}, among others \citep{planck_ame2015}, and potentially even in external galaxies \citep{Murphy_2010,Murphy2018,battistelli2019}. However, the polarized properties of this emission is still uncertain as physical models predict varying levels of potential polarized emission \citep{AME_state_of_play}. 

Ignoring this component during Galactic foreground removal can have a detrimental effect on polarized CMB determination. As demonstrated by \citet{Remazeilles_2016}, neglecting AME models with a polarization fraction as low as $\sim$\,1\,\% can cause a nonnegligible bias on the measured tensor-to-scalar ratio for a wide range of proposed future missions.

The currently favored model for AME is spinning dust emission, in which the radiation is due to small, rotating dust grains with an electric or magnetic dipole moment \citep{draine1998_2}. The total power $P_{\nu}$ emitted at frequency $\nu$ from a single grain rotating at frequency $\omega$ = $2\pi\nu$, electric or magnetic dipole moment $\mu$, and with angle $\theta$ between its dipole moment and rotation axis is given by
\begin{equation}
\centering
P_{\nu} = \frac{2}{3} \frac{\omega^4 \mu^2 \sin^2 \theta}{c^3} \, .
\label{eq:larmor}
\end{equation}
 Much of the theoretical work on spinning dust emission has been devoted to predicting the distribution of grain angular velocities present in the interstellar medium, which depends on properties such as grain size and shape as well as the local environment, such as the local gas density, temperature, and ionization \citep{draine1998_2,hoang2010,silsbee:2011}. These properties are poorly constrained, so theoretical predictions for the frequency spectrum of the AME are highly uncertain. 
 
The details of this model have been combined in the \texttt{SpDust2} code to produce AME spectra for varying interstellar medium conditions \citep{hoang2010,silsbee:2011}. In \cite{planck2014-a11}, the \commander\ \citep{eriksen2008} analysis had to combine two \texttt{SpDust2} spectra in order to get a good fit to the observed AME. While theoretical studies have attributed two component AME models to emission from the cold and warm neutral media \citep{hoang2010,ysard+}, a recent study found no link between the AME spectrum and the fraction of HI in the cold neutral phase \citep{Hensley21}, casting doubt on this interpretation. In this paper, we therefore seek constraints on AME properties without imposing strong assumptions on its frequency spectrum.

The polarization of spinning dust emission depends on how efficiently they align with the local magnetic field. \cite{hoang2013} argue that starlight polarization features in HD 197770 and HD 147933-4 could be caused by weakly aligned polycyclic aromatic hydrocarbons (PAHs), corresponding to polarization from spinning PAHs of $\lesssim 1~\%$ for frequencies $\sim$ 20~GHz. Additionally, \cite{Hoang_Lazarian_2016} calculated that iron nanoparticles can efficiently align with the Galactic magnetic field, yielding a wide range of potential polarization fractions of their microwave spinning dust emission. On the other hand, \cite{draine2016} argue that quantum suppression of energy dissipation interferes with alignment of very small grains, leading to extremely small polarization fractions at microwave frequencies.

Though the spinning dust model is favored due to intensity measurements, other models have not been definitively ruled out. A theory of emission from thermal fluctuations in magnetic dust grains has been proposed by \cite{draine1999}, and subsequently revisited by \cite{draine2013}. They identified resonance behavior that can produced peaked, AME-like spectra dependent on the magnetic properties and shapes of the grains. Predictions for polarization fractions in such resonances range from $\sim$5 to 40~\%, in conflict with current best limits on AME polarization. 

Theoretical predictions for the low-level polarized emission from AME are supported by most published observations. Using QUIJOTE data focused on molecular complex W43r, \cite{QUIJOTE_II_2016} found rigid upper limits for polarized AME at 16.7, 22.7, and 40.6\,GHz of 0.39, 0.52 and 0.22\,\% respectively. The most recent analysis of QUIJOTE data by \cite{QUIJOTE_III} concluded that the polarization fraction is indeed consistent with zero  within the Taurus molecular cloud. This analysis found the polarization fraction of AME to be less than $4.2\,\%$ ($95\,\%$ confidence), assuming that all of the observed polarized emission were to come from AME. \cite{battistelli2006} measured the total emission within the Perseus region to be polarized at $P = 3.4^{+1.5}_{-1.9} \,\%$ at 11\,GHz ($95\,\%$ confidence interval), while  \cite{planck2014-a31} derives a $2\,\sigma$ AME polarization percentage limit of less than $ 1.6\,\%$ within the same region. The differing results are likely due to the handling of polarized synchrotron emission, and this emphasizes the importance of synchrotron marginalization.

\cite{Macellari_2011} carried out a template based cross-correlation analysis with standard foreground templates to estimate the amplitude of the Galactic components in both intensity and polarization. Correlations were determined for synchrotron, dust, and free-free emission within the \WMAP\ 5-year maps using \cite{haslam1982}, the \WMAP\ 94\,GHz dust prediction, and the \cite{finkbeiner2003} H$\alpha$ maps as templates, respectively. With this method they found the polarization of both dust and free-free emission to be consistent with zero for the all-sky analysis. Concerning AME, this results in a full sky AME polarization fraction at \WMAP\ K-band of less than $5\,\%$ ($95\,\%$ confidence). They also highlight dust-correlated polarization fractions at a $2\,\sigma$ level within some regions of the sky within the K-band, while noting slightly negative dust polarizations (negative correlation between total intensity template and polarized emission) indicating a degeneracy between sky components.
 
In this work we add to earlier constraints in three key ways. Firstly, rather than investigating a specific region, we fit the level of polarized dust correlated emission using nearly the full sky. Secondly, we make no assumptions about the AME SED to avoid bias from an imperfect spectral fit. Instead, we make a strong spatial assumption, namely that AME is perfectly correlated (or anti-correlated) with thermal dust emission. Finally, we employ the Bayesian \BP\ framework \citep{bp01} to marginalize over the uncertainty in the other Galactic foregrounds and systematic effects. This approach attacks the CMB analysis problem by fitting instrumental, astrophysical and cosmological parameters all jointly within an integrated Gibbs sampling framework. The synchrotron spectral index and amplitude are independently redetermined within the current paper, to allow joint fits with the AME parameters. Technically speaking, marginalization over systematic uncertainties is implemented by analysing an ensemble of different \BP\ Low-Frequency Instrument (LFI) map realizations within the sampling scheme, thereby propagating low-level uncertainties from the \BP\ mapmaking procedure. This is the first example of how systematic errors may be propagated into postproduction analyses using Monte-Carlo ensembles with \lfi\ data.

The paper is organized as follows: Section~\ref{sec:data} describes the data used in this analysis. An overview of the sky model and sampling techniques are discussed in Sects.~\ref{sec:sky_model} and \ref{sec:algorithms} respectively, and in Sect.~\ref{sec:results} the results of the Gibbs sampler on the \BP\ data are discussed. Concluding remarks and outlook are reported in Sect.~\ref{sec:summary}. All software is made  publicly available.\footnote{\url{https://github.com/hermda02/dang}}

\section{Data}
\label{sec:data}

Because AME is only observed between 10-60~GHz in intensity, we restrict our data selection to the 20--70\,GHz frequency range. As such, our main observations are the polarization observations from the \Planck\ LFI \citep{planck2016-l02} and \WMAP\ \citep{bennett2012} experiments. For LFI, we adopt the \BP\ 30, 44, and 70\,GHz frequency sky maps\footnote{\url{http://beyondplanck.science}} \citep{bp01}, which come in the form of an ensemble of samples, each of which represents a different realization of systematic effects for the frequency channel in question. These include, but are not limited to, corrections for gain variations \citep{bp07}, correlated noise \citep{bp06}, bandpass leakage \citep{bp09} and far sidelobe contamination \citep{bp08}. In order to properly marginalize over these uncertainties, we draw maps from an ensemble of 1000 different map realizations. 

For \WMAP, we include the K- (22.8\,GHz), Ka- (33.0\,GHz), Q- (40.6\,GHz), and V-band (60.8\,GHz) channels. The W-band (94\,GHz) data are however excluded due to a high level of systematic residuals \citep{bennett2012}.

In addition to these main observations, we utilize the \Planck\ DR4 353-GHz polarization map \citep{planck2020-LVII} as a tracer of polarized dust emission. We use the raw DR4 353-GHz map, with no CMB subtraction, noting that this channel is strongly dust dominated. Finally, in order to derive an estimate of the polarization fraction $p_{\rm AME}$ of AME we need an estimate of the AME intensity. For this, we adopt the \Planck\ 2015 AME component map \citep{planck2014-a12}, rather than the corresponding \BP\ AME maps \citep{BP13}, as the former was derived without strong spatial priors.

All maps are smoothed to a common angular resolution of $1^{\circ}$ FWHM Gaussian beam, and discretized using the \healpix\footnote{\url{http://healpix.jpl.nasa.gov}} pixelization format \citep{gorski2005} with a resolution parameter of ${\nside = 64}$. We adopt the cosmological convention for the definition of the Stokes $Q$ and $U$ parameters, which differs from the IAU convention only in the sign of $U$ \citep{gorski2005}. Rayleigh-Jeans brightness temperature units ($\mu {\rm K_{RJ}}$) are utilized for the entirety of this work, and full bandpass integration for both \WMAP\ and \BP\ \lfi\ maps are taken into account during the fitting procedures.

To obtain a reliable estimate of the level of detectable polarized dust emission below 60\,GHz we need to mask portions of the sky which may bias our results. At the same time, it is highly desirable to include as much of the sky as possible, to maximize our signal-to-noise ratio. We therefore choose to primarily define our primary analysis mask on the central Galactic region defined in \citet{bp14}, but additionally we mask out a few compact regions with very high $\chi^2$'s, including Tau-A. A total of $f_{\mathrm{sky}}=91\,\%$ is included in the final analysis.

\section{Sky model}
\label{sec:sky_model}

\subsection{Modelling considerations}

In this paper, we make no assumptions regarding the AME spectral
energy density (SED). Instead, we note from the \Planck\ 2015 analysis
that AME in total intensity is highly correlated with thermal dust
emission \citep{planck2014-a12}.
This motivates us for now to assume that any detectable polarized AME
also has the same morphology as polarized thermal dust emission as
measured at 353\,GHz.  If the grains emitting the AME are aligned by
the same magnetic field aligning those responsible for FIR
polarization, and if the alignment is in the usual sense, that is the short
axis parallel to the magnetic field \textbf{B}, then the polarization
direction at AME frequencies is the same as that observed at 353
GHz \citep{draine1998_a}.

The polarization fraction of dust emission is a function of grain properties (alignment, shape, size), density structure, and magnetic field geometry. As shown in \cite{planck2016-l11B} using the \Planck\ full-sky maps, neither grain alignment nor the dust temperature drive variations in the polarization fraction over diffuse and translucent sightlines (for column densities $N_{\mathrm{H}} < 8\times10^{21}\,\mathrm{cm}^{-2}$, corresponding to 98~\% of the sky). The highest column densities where this conclusion breaks down lie, for the most part, within the portion of the sky which is masked out here. This conclusion is supported by \cite{Clark_2019}, who find that with \ion{H}{I} structures alone are able to reproduce the \Planck\ 353~GHz polarization maps at degree scales with high accuracy over most of the sky. Therefore, if the grains producing the AME and the far infrared continuum emission are aligned with the same magnetic field, we expect the polarization fraction of each to be similar. Nevertheless, it is possible that, unlike the larger grains, the alignment efficiency of nanoparticles varies substantially across the sky. In this case, the polarization fractions of the two emission mechanisms would be less correlated and AME polarization fractions in excess of those derived in this work would be permitted However, there is no strong observational evidence for this scenario and we do no consider it further in our analysis.

A known issue in AME polarization studies is the presence of polarized
synchrotron emission \citep{planck2014-a31}. Additionally, polarized
thermal dust emission is nonnegligible even at AME frequencies. In principle, we could disentangle these
distinct Galactic components by using a template for polarized
synchrotron emission in a way analogous to the dust emission. However,
no suitable high signal-to-noise ratio polarized synchrotron templates
currently exist. As an example, one might consider using the polarized
synchrotron component derived by the SMICA algorithm in
\cite{planck2016-l04}. However, this analysis explicitly assumes that
the polarized thermal dust emission ``vanishes at 30\,GHz'', and
therefore all polarized emission at 30\,GHz is assigned to be
exclusively synchrotron emission. This would lead to erroneously tight
constraints on polarized AME emission in the current analysis. In the absence of a high signal-to-noise ratio template of polarized
synchrotron emission, we adopt a simple parametric power-law model to
fit the synchrotron component in each pixel, and marginalize over the
free parameters.

\subsection{Parametric model}\label{sub:sky_model}

We model the observed emission ${d}_{\nu,p}$ in pixel $p$ at frequency $\nu$ as the sum of the true sky signal  ${s}_{\nu,\, p}$ and the noise ${n}_{\nu,\, p}$, given by
\begin{equation}
{d}_{\nu, p} = {s}_{\nu,p} + {n}_{\nu,p}.
\label{eq:data_model}
\end{equation}
The polarized sky signal, $s_{\nu,p}$, is usually described by three components in the microwave regime in addition to AME, namely thermal dust emission, synchrotron emission, and CMB \citep{planck2014-a31,planck2016-l04,bp14}. In this analysis, however, we do not include CMB because this contribution is known to be small over the frequency range and angular scales of interest \citep{planck2016-l05}, and including unconstrained degrees of freedom typically leads to nonphysical degeneracies. Instead, we perform dedicated sensitivity tests in Sect.~\ref{sub:cmb_test} by adding simulated CMB fluctuations to the existing sky maps, and show that this has a minor impact on final results. The sky model may therefore be written out in the following form,
\begin{equation}
{s}_{\nu,p} = {s}_{\mathrm{s},\nu,p} + {s}_{\mathrm{AME},\nu,p} + {s}_{\mathrm{d},\nu,p}. \label{eq:sky_simple} 
\end{equation}

Synchrotron emission is modeled  in terms of a power-law SED with a free amplitude per Stokes parameter and pixel, $a^{Q,U}_{\mathrm{s},p}$, and a spectral index, $\beta_{\mathrm{s},p}$, per pixel, but common for $Q$ and $U$,
\begin{equation}
s^{Q,U}_{\mathrm{s},\nu,p} =\, a^{Q,U}_{\mathrm{s},p} \,\Big(\frac{\nu}{\nu_{\rm 0, s}}\Big)^{\beta_{\mathrm{s},p}}.
\label{eq:synch}
\end{equation}
We choose a reference frequency of $\nu_{\rm 0, s}=30\,{\rm GHz}$ for synchrotron emission. 

For thermal dust emission, we adopt a single modified blackbody (MBB) SED,
\begin{equation}
{s}_{\mathrm{d},\nu,p}^{Q,U} = \mathcal{T}_{\mathrm{d},p}^{Q,U} \,\Big(\frac{\nu}{\nu_{0, \mathrm{d}}} \Big)^{\beta_{\rm d}+1}\, \frac{e^{h\nu_{0, \mathrm{d}}/kT_{\rm d}}-1}{e^{h\nu/kT_{\rm d}}-1}.\label{eq:thermal}
\end{equation}
where $h$ and $k$ are the Planck and Boltzmann constants, respectively; $\nu_{0, \mathrm{d}}$ is the thermal dust reference frequency, which is taken to be $353\,{\rm GHz}$; ${\beta_{\rm d}}$ is the thermal dust spectral index; and $T_{\rm d}=19.6\,{\rm K}$ is the dust temperature; and $\mathcal{T}_{\mathrm{d},\,p}$ is the \Planck\ 353\,GHz map. We marginalize over the thermal dust spectral index using the same prior as in the main \BP\ analysis ($\beta_{\rm d} = 1.62 \pm 0.04$; \citealp{bp14}) by drawing a new value for $\beta_{\rm d}$ in each Markov chain iteration. 

Finally, we assume that polarized AME is spatially perfectly correlated with polarized thermal dust emission, and we adopt the \Planck\ DR4 353\,GHz map as a spatial template for polarized AME. Allowing for a single multiplicative amplitude, $a_{\mathrm{AME},\, \nu}$, for both Stokes parameters, our model reads
\begin{align}
s_{\mathrm{AME},\, \nu,\, p}^{Q,U} &=  a_{\mathrm{AME},\, \nu}\, \mathcal{T}_{\mathrm{d},\,p}^{Q,U}.
\label{eq:AME}
\end{align}
However, to avoid a
perfect degeneracy between the spatial synchrotron template and the
AME template amplitudes, a maximum of $N_\mathrm{band}-1$ template
amplitudes can be fit. To break this degeneracy, we fix
$a_{\mathrm{AME},\, \nu}=0$ for the \WMAP\ V-band (61\,GHz) and LFI
70\,GHz channels, where the AME-to-thermal-dust ratio in intensity is
$0.033$ and $0.008$, respectively, in the \Planck\ 2015 model
\citep{planck2014-a12}.

Collecting terms, we arrive at the final complete data model,
\begin{equation}
s_{\nu,p}^{Q,U} = a_{\mathrm{s},p}^{Q,U} \,\left(\frac{\nu}{\nu_{\rm 0, s}}\right)^{\beta_{\mathrm{s},p}} + \mathcal{T}_{\mathrm{d},p}^{Q,U}\left(\left(\frac{\nu}{\nu_{0, \mathrm{d}}} \right)^{\beta_{\mathrm{d}}+1}\, \frac{e^{h\nu_{0, \mathrm{d}}/kT_{\rm d}}\text{-}1}{e^{h\nu/kT_{\rm d}}\text{-}1} + a_{\mathrm{AME}, \nu} \right)\label{eq:data_model_full}
\end{equation}
In this model we fit $a_{\mathrm{s},p}^{Q,U}$ and $\beta_{\mathrm{s},p}$ per pixel, and $a_{\rm AME,\nu}$ as a single value across the sky. The synchrotron amplitude is determined independently in Stokes $Q$ and $U$, while both $\beta_{\mathrm{s}}$ and $a_{\rm AME,\nu}$ are fit jointly in $Q$ and $U$. The primary goal of the following analysis is to constrain the $a_{\rm AME,\nu}$ coefficients, and the other parameters are simply considered to be nuisance parameters that we marginalize over.

\subsection{From AME template amplitudes to polarization fraction}\label{sub:ame}

The AME polarization fraction $p_{\mathrm{AME}}$ is defined as the ratio of the polarized ($P_{\mathrm{AME},\,\nu,\,p}$) and total AME intensities ($I_{\mathrm{AME},\,\nu,\,p}$) at each frequency $\nu$ and pixel $p$, i.e. $P_{\mathrm{AME},\,\nu,\,p}/I_{\mathrm{AME},\,\nu,\,p}$.

To translate the vector of template amplitudes, $\boldsymbol a_{\rm AME}$, to an AME polarization fraction, $ p_{\rm AME,\,\nu}$, we use Eq.~\eqref{eq:AME} to associate polarized AME with polarized thermal dust emission,

\begin{equation}
P_{\mathrm{AME},\,\nu,\,p}=a_{\mathrm{AME},\,\nu}\sqrt{{\mathcal{T}_{\mathrm{d},\,p}^{Q}}^2+{\mathcal{T}_{\mathrm{d},\,p}^{U}}^2}=a_{\mathrm{AME},\,\nu}~P_{353,p}.
\end{equation}
Following the results from \Planck\ 2015, we can describe $I_{\mathrm{AME},\,\nu,\,p}$ as the product of the frequency dependence $f_{\mathrm{AME}}(\nu)$ and the AME amplitude map $A_{\mathrm{AME},\,p}$, yielding

\begin{align}
  p_{\mathrm{AME},\,\nu,\,p} &= a_{\mathrm{AME},\, \nu}\, P_{353,\,p} / I_{\mathrm{AME},\,\nu,\,p}\\
                 &= p_{353,\,p}\frac{a_{\mathrm{AME},\, \nu}}{f_{\mathrm{AME}}(\nu)} \frac{I_{353,\,p}}{A_{\mathrm{AME},\,p}},
\label{eq:pfrac_def2}
\end{align}
where $p_{353,\,p}$ is equivalently defined as the polarization fraction at 353~GHz, i.e. $p_{353,\,p}=P_{353,\,p}/I_{353,\,p}$. We estimate the maximum AME polarization fraction $p_{\mathrm{AME},\,\nu}^{\mathrm{max}}$, corresponding to the magnetic field in the plane of the sky as
\begin{equation}
p_{\mathrm{AME},\,\nu}^{\mathrm{max}} = p_{353}^{\mathrm{max}}\frac{a_{\mathrm{AME},\, \nu}}{f_{\mathrm{AME}}(\nu)} \large\langle \frac{I_{353}}{A_{\mathrm{AME}}}\large\rangle,
\label{eq:pfrac_def3}
\end{equation}
 where $p_{353}^{\rm max}$ is the maximum dust polarization fraction at 353~GHz and $\large\langle \frac{I_{353}}{A_{\mathrm{AME}}}\large\rangle$ is the ratio of the 353 GHz dust emission to 22.8~GHz AME emission, assumed to be constant over the sky in analogy with our template analysis for polarization. We adopt $p_{353}^{\rm max} = 22\,\%$ \citep{planck2016-l11B}, and $\langle I_{\rm 353}/A_{\rm
   AME}\rangle\,=\,2.5^{+0.2}_{-0.7}$ 
 and $f_{\mathrm{AME}}(\nu)$ are taken from the \Planck\ 2015 data products \citep{planck2014-a12}. This allows us to estimate $p_{\mathrm{AME},\,\nu}^{\mathrm{max}}$ directly from our fit amplitudes $a_{\mathrm{AME},\, \nu}$.

We note that Eq.~\eqref{eq:pfrac_def2} provides a direct linear
estimate of the AME polarization fraction in terms of the fitted AME
template amplitudes, which themselves are fitted linearly to the raw
data. As such, Eq.~\eqref{eq:pfrac_def2} supports simple Gaussian error
propagation without explicit noise debiasing, which otherwise often is
a problem for polarization fraction estimates with low signal-to-noise
data; readers can refer to \cite{pma2}, for example. With the current
formulation, this issue is circumvented through linear association
with high signal-to-noise thermal dust emission estimates that themselves have
been noise debiased.

\section{Sampling algorithms}\label{sec:algorithms}
\subsection{Posterior distribution and Gibbs sampling}\label{sub:gibbs}

Following the general approach in the \BP\ environment, we employ
Bayesian sampling methods to estimate the various free parameters in our
model. Let us define the set of all free parameters in
Eq.~\eqref{eq:data_model_full} as $\vec{\omega} = \{\a_{\mathrm{AME}},
\a_{\mathrm{s}}, \beta_{\mathrm{s}}, \beta_{\mathrm{d}}\}$. The
appropriate posterior distribution is then given by Bayes' theorem,
\begin{equation}
P(\vec{\omega}\mid\boldsymbol{d}) = \frac{P(\boldsymbol{d} \mid \vec{\omega}) P(\vec{\omega})}{P(\boldsymbol{d})} \propto \mathcal{L}(\vec{\omega}) P(\vec{\omega}),
\label{eq:bayes}
\end{equation}
where $ P(\boldsymbol{d}\mid\vec{\omega}) \equiv \mathcal{L}(\vec{\omega})$ is the likelihood function, $P(\vec{\omega})$ is the prior, and $P(\boldsymbol{d})$ is a normalization factor which we disregard here as it is independent of the parameters $\vec{\omega}$.

\begin{figure}[t]
\centering
\includegraphics[width=0.475\linewidth]{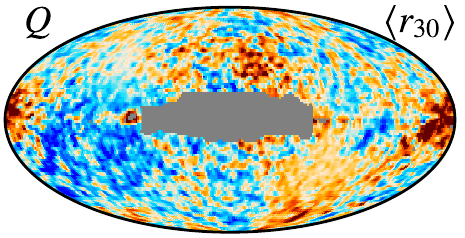}
\includegraphics[width=0.475\linewidth]{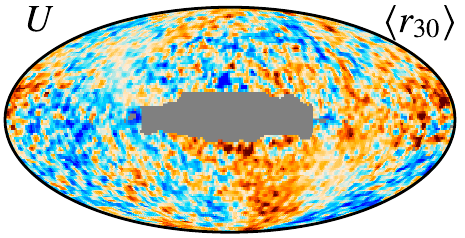}
\includegraphics[width=0.475\linewidth]{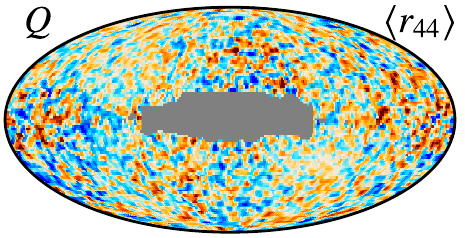}
\includegraphics[width=0.475\linewidth]{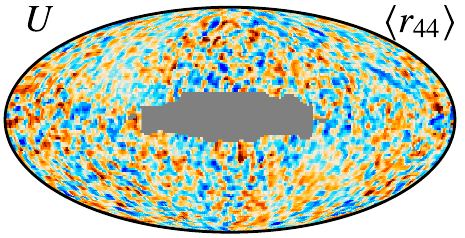}
\includegraphics[width=0.475\linewidth]{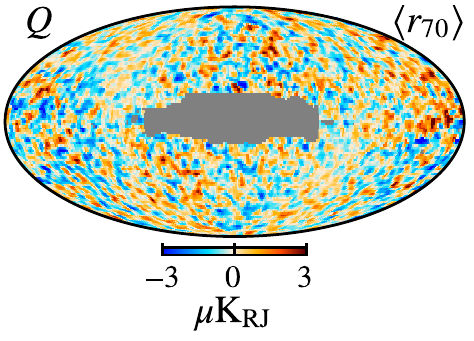}
\includegraphics[width=0.475\linewidth]{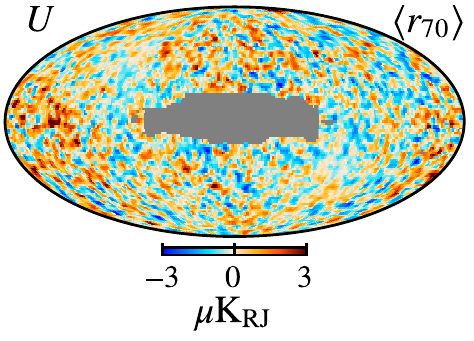}
\includegraphics[width=0.475\linewidth]{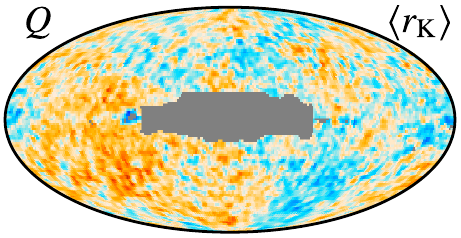}
\includegraphics[width=0.475\linewidth]{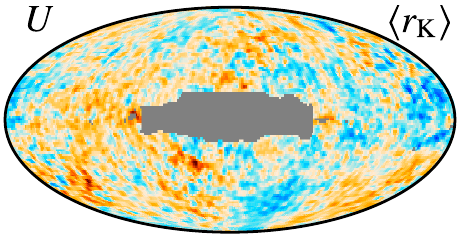}
\includegraphics[width=0.475\linewidth]{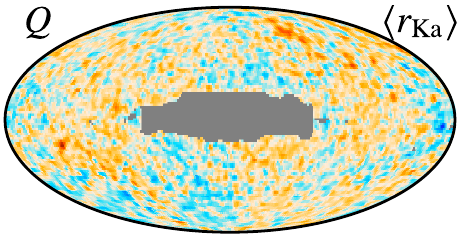}
\includegraphics[width=0.475\linewidth]{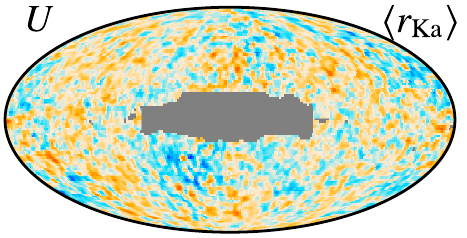}
\includegraphics[width=0.475\linewidth]{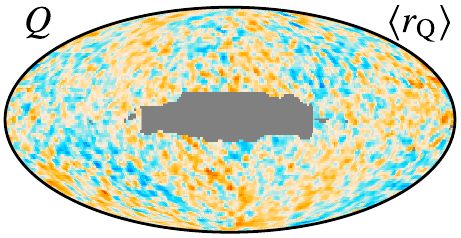}
\includegraphics[width=0.475\linewidth]{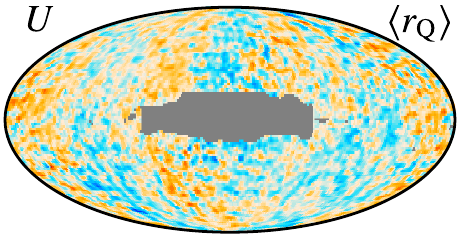}
\includegraphics[width=0.475\linewidth]{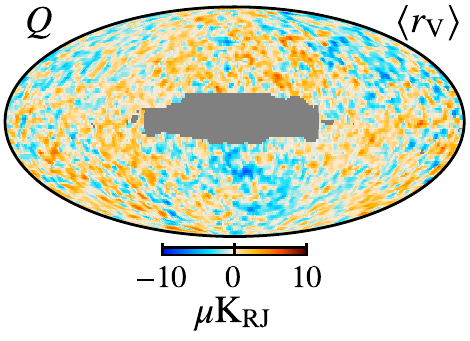}
\includegraphics[width=0.475\linewidth]{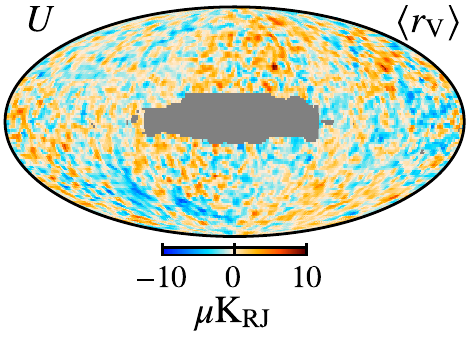}
\caption{Posterior mean residual maps
  ($\boldsymbol{r}_{\nu}\,=\,\boldsymbol{d}_{\nu}\,-\,\boldsymbol{s}_{\nu}$)
  in Stokes $Q$ and $U$ for (from top to bottom) the \BP\ LFI 30, 44,
  and 70\,GHz maps and the \WMAP\ K, Ka, Q, and V bands. All maps are
  smoothed to a common resolution of $3^\circ$ FWHM.}
\label{fig:res}
\end{figure}

\begin{figure}[t]
\centering
\includegraphics[width=\linewidth]{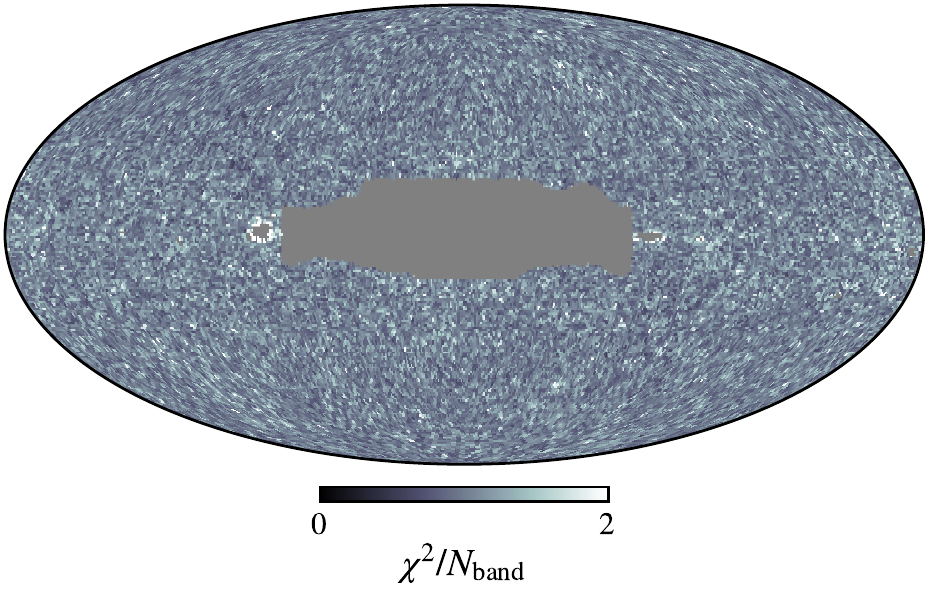}
\caption{Mean $\chi^2/N_{\mathrm{band}}$ of all Gibbs iterations, averaged over Stokes $Q$ and $U$.}
\label{fig:chisq}
\end{figure}

\begin{figure}[t]
\centering
\includegraphics[width=\linewidth]{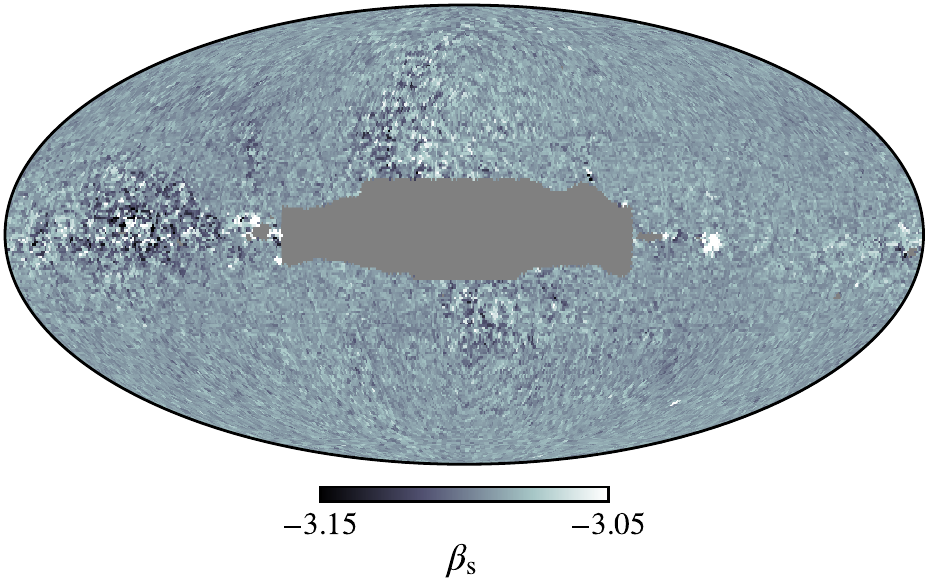}
\caption{Posterior mean polarized synchrotron spectral index map.}
\label{fig:synch_beta}
\end{figure}

In order to map out the probability distribution of our
multidimensional parameter space in a computationally efficient way,
we adopt Gibbs sampling \citep[e.g.,][]{Gelman03}. In the Gibbs
sampling framework we cycle through each parameter independently,
drawing each parameter from a distribution conditional on the values
of the other parameters; see, e.g., \citet{bp01} for details of the
much larger \BP\ Gibbs sampler. In the present work, the Gibbs sampler
takes the following form,
\begin{align}
(\a_{\mathrm{AME}},\a_{\mathrm{s}}) &\leftarrow P(a_{\mathrm{AME}}, \a_{\mathrm{s}}|
  \d, \beta_{\mathrm{s}}, \beta_{\mathrm{d}})\label{eq:amp_samp}\\
\beta_{\mathrm{s}}  &\leftarrow P(\beta_{\mathrm{s}}|
\d, \a_{\mathrm{AME}}, \a_{\mathrm{s}}, \beta_{\mathrm{d}}) \label{eq:beta_samp}\\
\beta_{\mathrm{d}}  &\leftarrow P(\beta_{\mathrm{d}}).
\label{eq:gibbs}
\end{align}
The two first steps will be detailed below, while the third step just
denotes sampling the thermal dust spectral index from its Gaussian
prior, as discussed in Sect.~\ref{sub:sky_model}.

\begin{figure*}[p]
\centering
\includegraphics[width=0.49\linewidth]{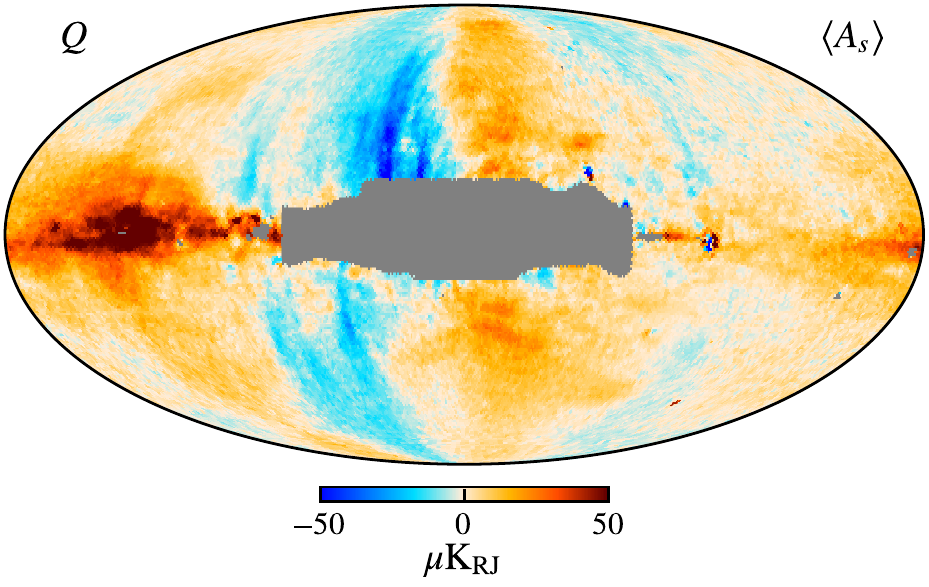}
\includegraphics[width=0.49\linewidth]{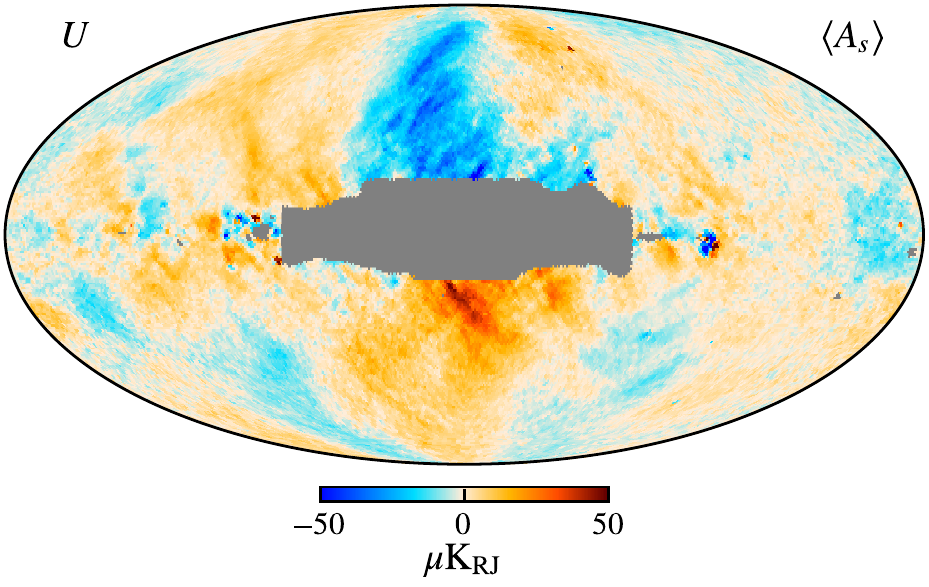}\\
\includegraphics[width=0.49\linewidth]{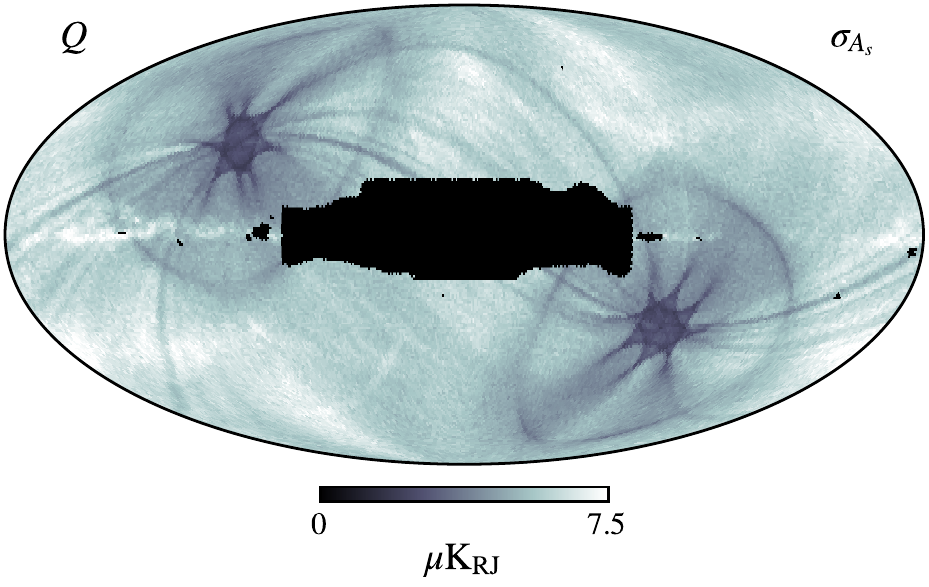}
\includegraphics[width=0.49\linewidth]{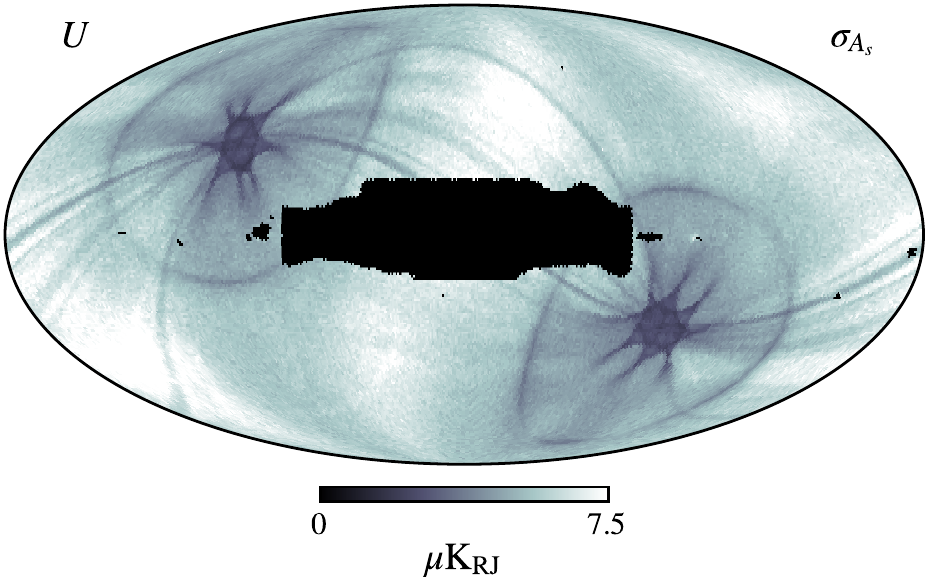}
\caption{Posterior mean (\textit{top}) and standard deviation (\textit{bottom}) maps of the synchrotron amplitude at ${\rm 30\,GHz}$ in Stokes Q (\textit{left}) and U (\textit{right}).}
\label{fig:synch}
\vspace*{5mm}
\includegraphics[width=0.49\linewidth]{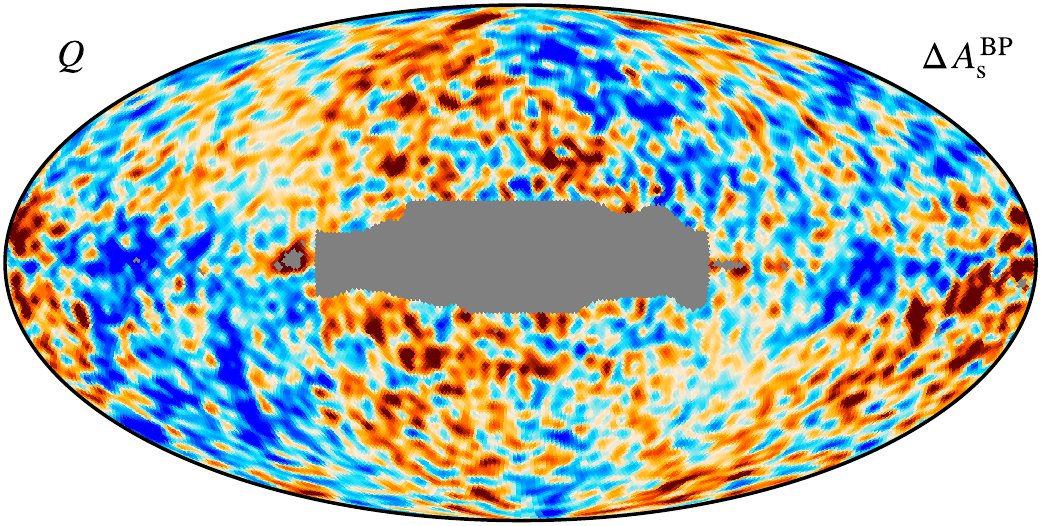}
\includegraphics[width=0.49\linewidth]{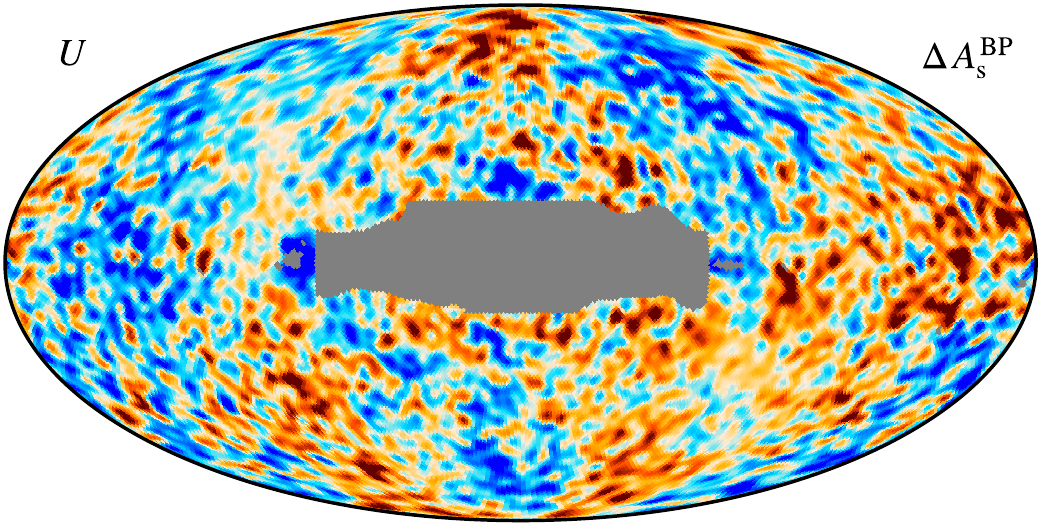}\\
\includegraphics[width=0.49\linewidth]{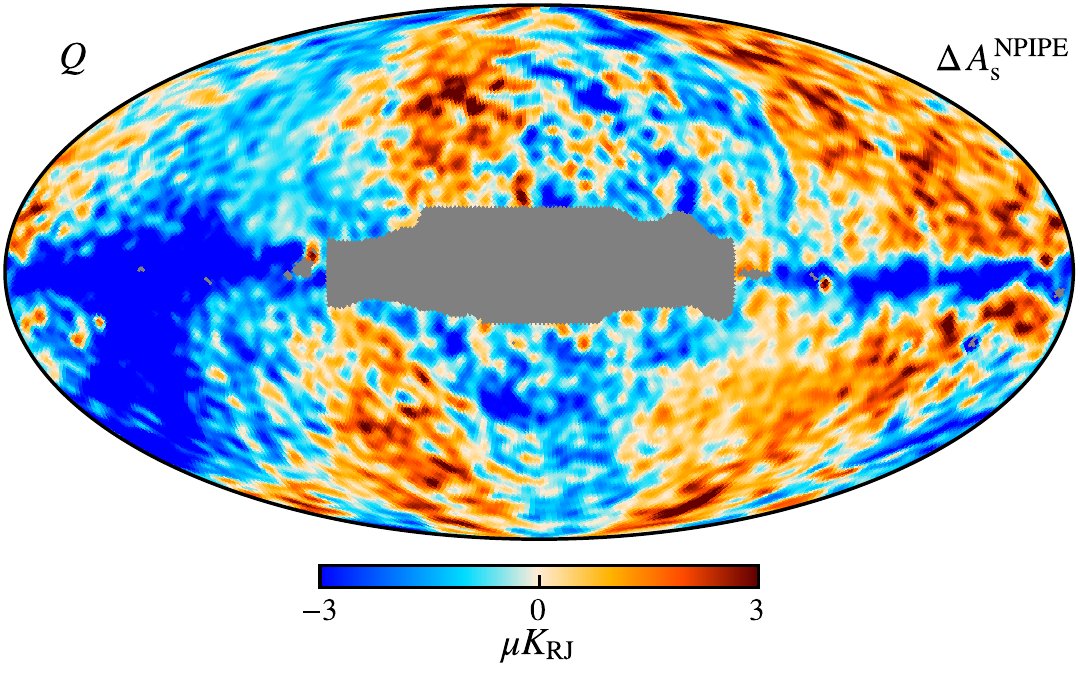}
\includegraphics[width=0.49\linewidth]{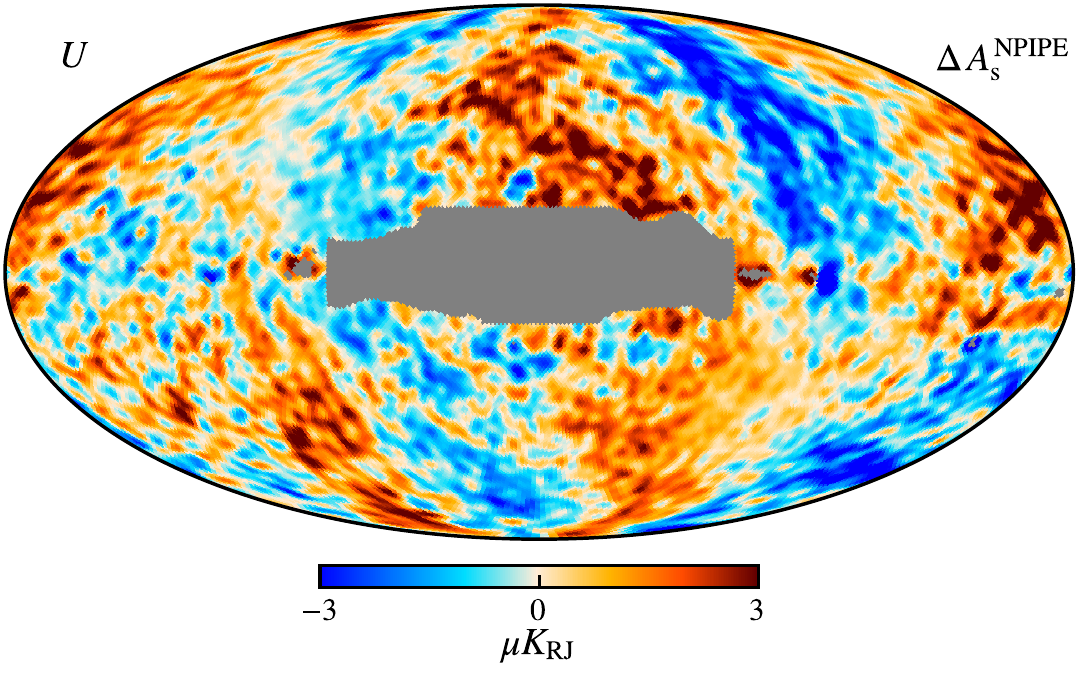}
\caption{Difference between the synchrotron solution derived here and the solution derived in other works. (\textit{Top}) Difference between the \BP\ synchrotron solution \citep{bp14} and the synchrotron component derived here. (\textit{Bottom}) Difference between \Planck\ DR4 and the synchrotron component derived here. We note a difference along the Galactic plane, which is likely due to the strong degeneracy between dust and synchrotron emissions in this region. The left and right columns correspond to Stokes $Q$ and $U$, smoothed with a $3^\circ$ FWHM Gaussian beam.}
\label{fig:BP_npipe_diff}
\end{figure*}

\subsection{Amplitude sampling}\label{sub:joint}

To sample from the conditional distribution for the amplitudes in
Eq.~\eqref{eq:amp_samp}, we note from Eq.~\eqref{eq:data_model_full}
that the signal model is linear in both $\a_{\mathrm{AME}}$ and
$\a_{\mathrm{s}}$. The data model in Eq.~\eqref{eq:data_model} may
therefore be written in the following compact vector form,
\begin{equation}
{d}_{\nu, p} = \tens{T}_{\nu}\vec{a} + s_{\mathrm{d},\nu} + {n}_{\nu,p},
\end{equation}
where
{\tiny
\begin{align}
\vec{a} &\equiv\begin{bmatrix}
a_{\rm s, 0} & \cdots & a_{\rm s, N_{\rm pix}-1} & & a_{\rm AME, 1} & \cdots & a_{\rm AME, N_{\rm bands} - 1} 
\end{bmatrix}^T\nonumber\\
\tens{T}_{\nu} &\equiv\begin{bmatrix}
(\frac{\nu}{\nu_{0,\,{\rm s}}})^{\beta_{{\rm s},0}} & 0 & 0 & \delta_{\nu,1}\mathcal{T}_{\mathrm{d},\,0}&\cdots& \delta_{\nu,n-1}\mathcal{T}_{\mathrm{d},\,0}\\
 0 & \ddots & 0 & \vdots & \cdots & \vdots\\
 0 & 0 & (\frac{\nu}{\nu_{0,\,{\rm s}}})^{\beta_{{\rm s},N_{\rm pix}-1}} & \delta_{\nu,1}\mathcal{T}_{\mathrm{d},\,N_{\rm pix}-1}&\cdots& \delta_{\nu,n-1}\mathcal{T}_{\mathrm{d},\,N_{\rm pix}-1}
\end{bmatrix}.\nonumber
\end{align}}
In these expressions, indices for Stokes $Q$ and $U$ parameters are
suppressed, but we note that all frequency or component maps should be
considered as stacked vectors, $\a = \{\a^Q,\a^U\}$, and matrices are
generalized accordingly.

With this notation, $\d_{\nu} - \T_{\nu}\a -
\s_{\mathrm{d},\nu} = \n_{\nu}$, which we assume to be Gaussian
distributed with a covariance matrix $\N_{\nu}$. Therefore, the
likelihood may be written in terms of a standard multivariate Gaussian,
\begin{equation}
  \mathcal{L}(\omega) \propto e^{-\frac{1}{2}\sum_{\nu}(\d_{\nu} - \T_{\nu}\a -
    \s_{\mathrm{d},\nu})^t\N_{\nu}^{-1}(\d_{\nu} - \T_{\nu}\a -
    \s_{\mathrm{d},\nu})} \equiv e^{-\frac{1}{2}\chi^2(\a,\beta)}
  \label{eq:likelihood}
\end{equation}
When interpreted as a function of $\a$ only, this is a strict Gaussian
distribution from which a sample may be drawn through the following
set of linear equations (see Appendix~A of \citealp{bp01} for a pedagogical
derivation),
\begin{equation}
\sum_{\nu}\left(\tens T_{\nu,\,j}^{T}\tens  N^{-1}_{\nu}\tens  T_{\nu,\,j}\right)\boldsymbol a=\sum_{\nu}\left(\tens T_{\nu,\,j}^{T} \tens N_\nu^{-1}\boldsymbol (\d_{\nu}-\s_{\mathrm{d},\nu}) + \tens T_{\nu,\,j}^T \tens N_\nu^{-1/2} \boldsymbol\eta_{j} \right).\nonumber
\label{eq:joint_lin_sample}
\end{equation}
Here $\boldsymbol\eta_{j}$ is a vector of standard Gaussian random
variates, $N(0,1)$. This equation may be solved efficiently through standard
Conjugate Gradient techniques \citep{shewchuk:1994}.

\subsection{Sampling the synchrotron spectral index}\label{sub:metrop}

Sampling the synchrotron spectral index has been an active topic of analysis throughout the history of both \textit{WMAP} and \Planck\ . Both the \textit{WMAP} and \Planck\ Collaborations note difficulties in independently determining $\beta_{\mathrm{s}}$ in polarization and therefore adopted the determination  from measurements in total intensity \citep{wmap_nine_year_final,planck2016-l04}. \cite{dunkley_2009} assumed a prior of $\beta_{\mathrm{s}}=-3.0\pm0.3$ from \cite{rybicki_lightman}, finding a mean index of $-3.03\pm0.04$ in regions of the sky with high signal-to-noise. More recent analyses including S-PASS data find spatial variations in the spectral index ranging between $\beta_{\mathrm{s}}\sim -2.8$ at low Galactic latitudes, and $\beta_{\mathrm{s}}\sim -3.1$ at higher Galactic latitudes, though both are consistent with $\beta_{\mathrm{s}}\sim -3.1$. Additionally, recent work with C-BASS data shows the average synchrotron spectral index between 4.76 and 22.8 GHz to be $\beta_{\mathrm{s}} = -3.1 \pm 0.02$ \citep{CBASS_gal_em}. Here we explore a variety of priors for with $-3.3 < \langle \beta_{\mathrm{s}} \rangle < -3.0$, and assume a nominal prior of $\beta_{\mathrm{s}} = -3.1\pm0.1$, following the results of \cite{bp14}.

To sample from Eq.~\eqref{eq:beta_samp}, we employ a Metropolis Markov
Chain Monte Carlo sampler \citep{metropolis:1953,bp01}. Specifically,
let $\beta_{j}$ represent the $j$th sample of $\beta_{\rm s}$ in a
Markov chain (implicitly suppressing the synchrotron label for
readability), and let $T(\beta_{j+1} |
\beta_{j})$ be a symmetric stochastic proposal probability
distribution, i.e., $T(\beta_{j+1} | \beta_{j})
= T(\beta_{j} | \beta_{j+1})$; we adopt a simple
Gaussian distribution with a mean of $\beta_{j}$ and a
(tunable) standard deviation of $\sigma$.

The Metropolis algorithm is then given by the following steps: First, the chain is initialized at some parameter value $\beta_0$, which can chosen as a good estimate to the parameter $\beta$ is as some random value. Secondly, a random sample is drawn from the proposal distribution, as $\beta_{j+1} \leftarrow  T(\beta_{j+1} \mid \beta_j)$, and  define $\omega_j = \{\a_{\mathrm{AME}}, \a_{\mathrm{s}}, \beta_{j}, \beta_{\mathrm{d}}\}$, while keeping all other parameters fixed. In order to decide if the sample should be accepted or rejected, the Metropolis acceptance probability $q$ is computed,
\begin{equation}
q = \frac{\mathcal{L}(\omega_{j+1})}{\mathcal{L}(\omega_{j})}\frac{P(\omega_{j+1})}{P(\omega_{j})},
\label{eq:accept_prob}
\end{equation}
where $\mathcal{L}(\omega)$ is given by Eq.~\eqref{eq:likelihood} and $P(\omega)$ denotes a Gaussian prior. Once $q$ is calculated, a random number, $\eta$, is drawn from a uniform distribution $U[0,1]$, and the proposed $\omega_{j+1}$ is accepted if $\eta<q$. If the sample is rejected, set $\omega_{j+1} = \omega_j$. This procedure is repeated until convergence.

An intuitive observation regarding the Metropolis method is that all
samples with a higher posterior probability are accepted, while some
samples with a lower posterior probability are accepted. By accepting
with a probability given by the ratio of the sample posteriors, the
density of samples within a given parameter volume is proportional to
the underlying probability density. In this work the proposal
probability distribution is Gaussian with a width $\sigma$ determined
by the standard deviation of the Gaussian prior distribution. An
ensemble of Gaussian priors is used in this analysis to explore the
effect of the prior on the final results.

\section{Results}
\label{sec:results}

We are now ready to present the results derived by applying the Gibbs
sampler described in Sect.~\ref{sec:algorithms} to the data summarized
in Sect.~\ref{sec:data}. Unless otherwise noted, the default synchrotron spectral index prior is $\beta_{\mathrm{s}}=-3.1\pm0.1$.
For each chain, 500 full Gibbs
samples are generated. Between each main sample, we recall that the
LFI frequency maps are replaced by different \BP\ samples to
marginalize over systematic effects, as discussed in
Sect.~\ref{sec:data}.

\subsection{Goodness-of-fit}\label{subsub:goodness}

Before examining the details of each emission component, we consider
the overall goodness-of-fit of our data model. Figure~\ref{fig:res}
shows mean residual maps for each frequency map on the form $\r_{\nu}
= \d_{\nu} - \T_{\nu}\a - \s_{\mathrm{d},\nu}$ for both LFI and
\WMAP. The gray region indicates the analysis mask discussed in
Sect.~\ref{sec:data}. Generally we observe an excellent fit in terms
of Galactic signal suppression; the Galactic plane is only barely
visible at the level of a few microkelvin in the LFI 30\,GHz and
\WMAP\ K-band. However, at higher latitudes there are clear
large-scale residuals at the level of $\lesssim 3\,\mu {\rm K}$ for
LFI and $\lesssim 5\,\mu {\rm K}$ for \WMAP\ with a clear
instrumental origin; for a detailed discussion of these structures, we
refer the interested reader to \citet{jarosik2007}, \citet{bp01}, \citet{bp14}, and \citet{bp17}.

It is worth noting that the LFI residual maps shown in
Fig.~\ref{fig:res} appear significantly different from the
corresponding maps in the main \BP\ analysis. The reason for this is,
as discussed earlier, that the \WMAP\ K-band channel is excluded from
the latter analysis. And from this plot we can intuitively understand
why: The high signal-to-noise of the \WMAP\ K-band heavily influences
the fit of the \Planck\ 30\,GHz band, and potential uncontrolled
systematic effects in the K-band may therefore compromise the LFI data
themselves. We also note that the 30\,GHz and K-band residuals are
morphologically similar, but with opposite signs, indicating
good agreement in terms of the Galactic signal, but with significantly
different large-scale scanning-induced systematics. For the main \BP\ analysis, which focuses on
LFI, it is therefore clear why the K-band is omitted---while in the
current paper, we want to establish a joint estimate between LFI and
\WMAP, and we want to maximize our signal-to-noise ratio, and we
therefore include all data. 

Figure~\ref{fig:chisq} shows a corresponding $\chi^2$ map, which has been normalized by the number of bands. At this level, we observe very little discernible structure, except for small excesses very near the mask, indicating that the adopted mask indeed performs well.

\subsection{Synchrotron component}
\label{sec:synch}

Next we examine the derived synchrotron posterior component maps.
Figure~\ref{fig:synch_beta} shows the posterior mean map for the
synchrotron spectral index, and we recall that the prior in this case
is $\beta_{\mathrm{s}}=-3.1\pm0.1$. Overall we see that $\beta_{\rm
  s}$ is very close to the prior mean over large portions of the sky,
though we can clearly see the impact from the likelihood in some high
signal-to-noise regions of the sky. In particular, we note significant
deviations from the prior along the Northern Galactic Spur, in the Fan
Region, and along the Galactic Plane, in agreement with
\citet{bp14}. However, by far the most important conclusion to be
drawn from this figure is the fact that even the combination of LFI
and \WMAP\ has very little constraining power compared to the prior with respect to the
spectral index of synchrotron emission, and all results that depends
on this will necessarily be sensitive to the assumed prior. In the
following, we therefore consider the synchrotron index prior mean
as a free hyperparameter, and all main results will be plotted as a
function of this parameter.

Next, the synchrotron Stokes $Q$ and $U$ amplitude components are
summarized in the form of posterior mean (top panels) and standard
deviation (bottom panels) maps in
Fig.~\ref{fig:synch}. Figure~\ref{fig:BP_npipe_diff} shows difference
maps between the synchrotron amplitude map derived in this paper and
those generated by the \BP\ (top row) and \Planck\ DR4 (bottom row)
pipelines. We find excellent agreement with the different pipelines to around the $3~\muK_{\mathrm{RJ}}$ level. The most dominant differences are systematic in nature. In
particular, we note that the main difference between the \BP\ and
current analysis is the addition of \WMAP\ K-band, while the
difference between the current analysis and the \Planck\ DR4
additionally highlights the difference between the end-to-end Bayesian
global approach \citep{bp01} and the classic frequentist
single-channel approach \citep{npipe}.

\begin{figure}[t]
\centering
\includegraphics[width=\linewidth]{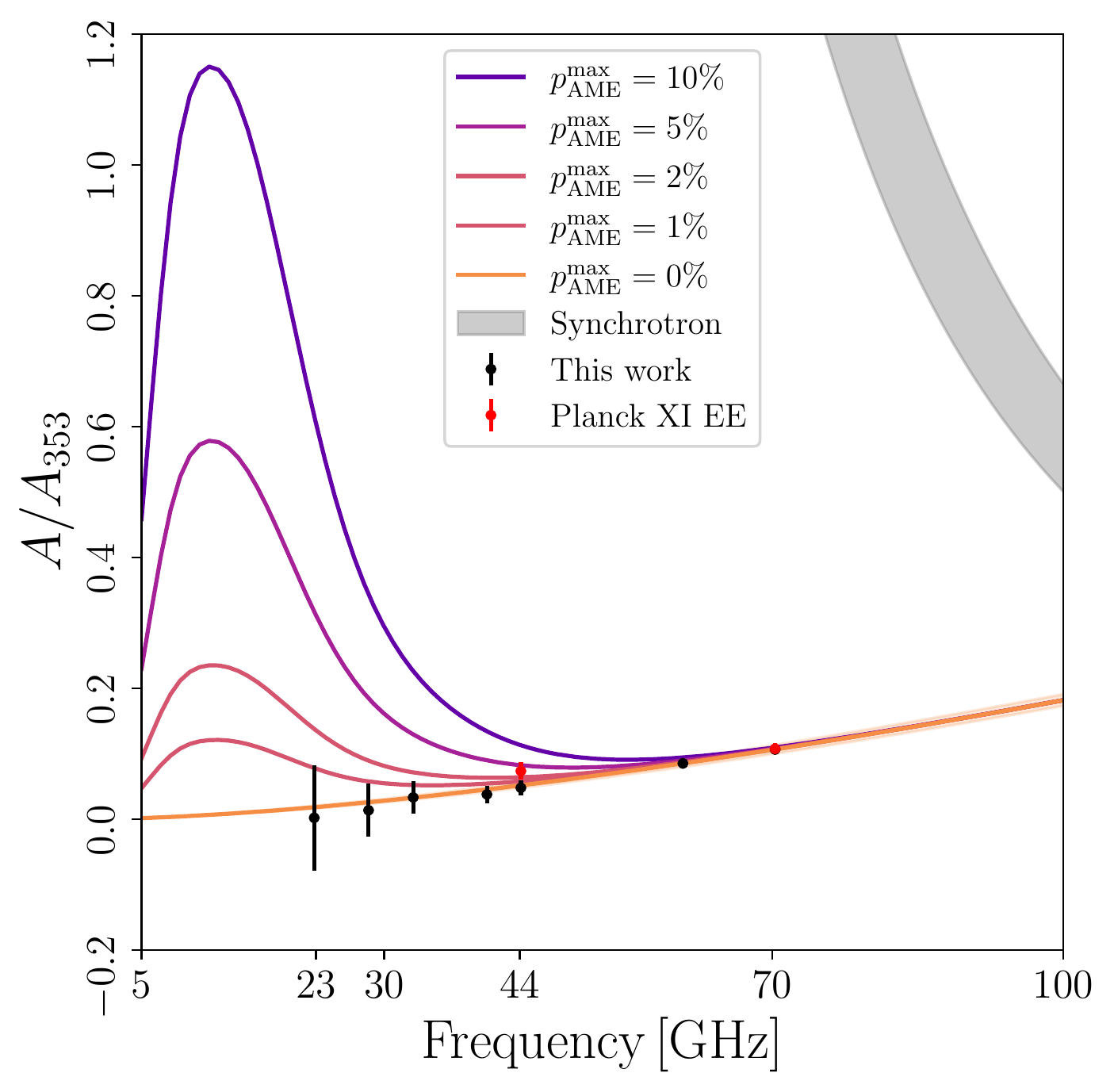}
\caption{Fitted AME amplitudes for a synchrotron prior of $\beta{\rm s}\sim N(-3.1,0.1)$, with various models for different levels of $p_{\rm AME}^{\mathrm{max}}$ overplotted as colored curves. The red points are taken from \citet{planck2016-l11A}.}
\label{fig:ame_seds}
\end{figure}

\begin{figure}[t]
\includegraphics[width=\linewidth,right]{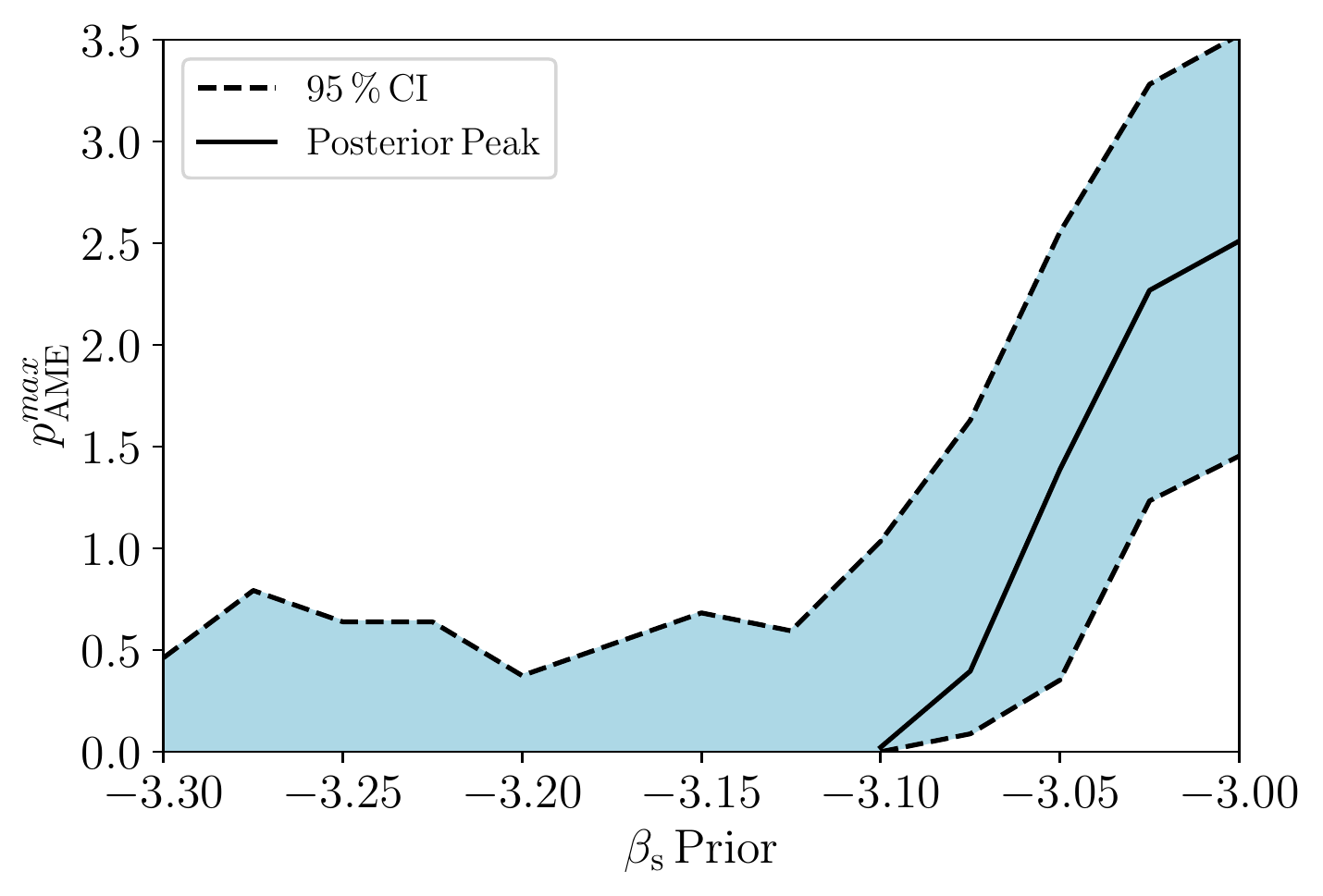}
\caption{$95\,\%$ confidence intervals for $p_{\rm AME}^{\rm max}$ as a function of the $\beta_{\rm s}$ prior mean.
}
\label{fig:pmax_ci}
\end{figure}

\begin{figure}[t]
\includegraphics[width=\linewidth,right]{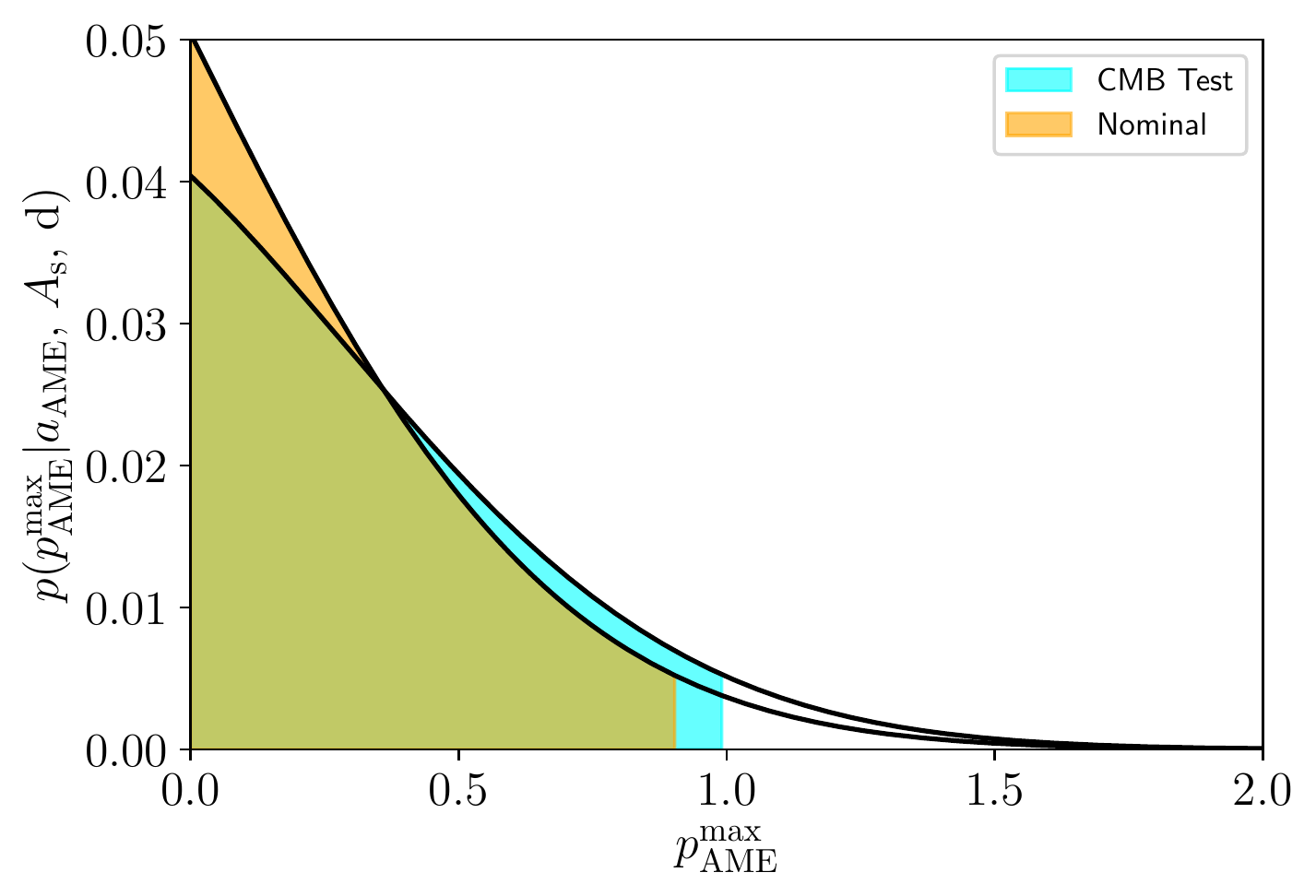}
\caption{Comparison of likelihood slices for $p_{\rm AME}^{\rm max}$ for a synchrotron prior of $\beta_{\rm s}\,\sim\,N(-3.1,0.1)$ with (blue) and without (orange) additional CMB fluctuations.}
\label{fig:posteriors}
\end{figure}

\begin{figure}[t]
\centering
\includegraphics[width=\linewidth]{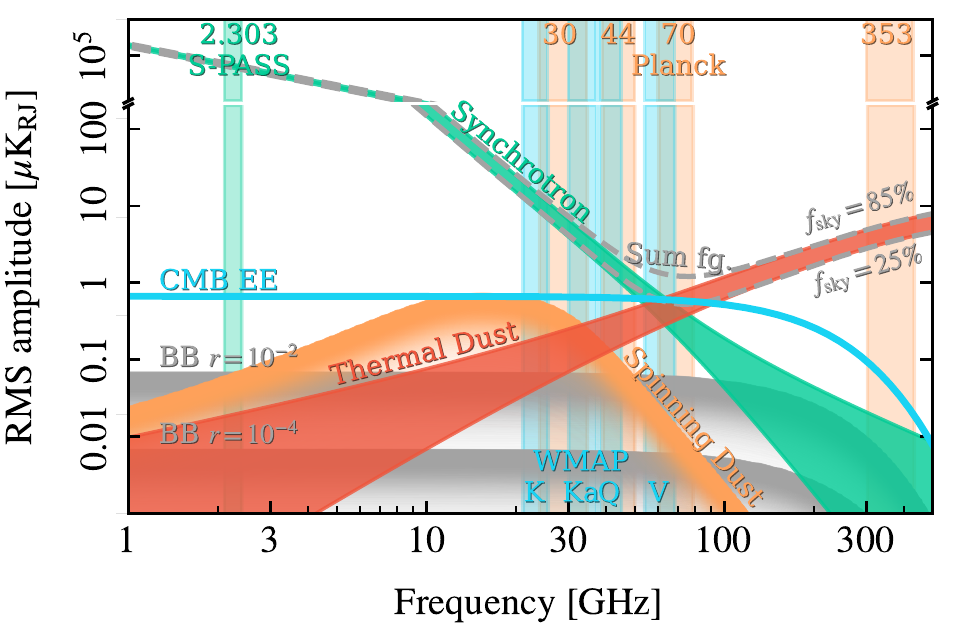}
\caption{Polarization spectrum in the microwave frequency range. In addition to the traditional spectrum which only includes thermal dust and synchrotron emission, an AME (spinning dust) component is added here with a maximum polarization fraction of $1\,\%$. The thermal dust and synchrotron spectra are determined by results from \BP\ \lfi\ analysis \citep{bp14}. The cyan and orange vertical bands correspond to the \WMAP\, and \Planck\ bands respectively.}
\label{fig:nuref}
\end{figure}

\subsection{Constraints on AME polarization}\label{sec:AME_constraints}

Finally, we are ready to present the main results of the current
paper, namely constraints on polarized AME emission, and we start the
linear amplitude parameters, $\a_{\mathrm{AME}}$. As an example, these
are shown in Fig.~\ref{fig:ame_seds} for a synchrotron prior of
$\beta_{\mathrm{s}}=-3.1\pm0.1$. For convenience, we normalize
$\a_{\mathrm{AME}}$ with respect to the 353\,GHz channel, and we plot
the sum of the thermal dust emission and AME, with different AME
polarization fractions marked as colored curves, ranging between 0 and
10\,\%.

As discussed in Sect.~\ref{sub:sky_model}, AME amplitudes are not fit at
\lfi\ 70\,GHz and \WMAP\ V-band as these bands are assumed to be fully
described by synchrotron and thermal dust emission. However,
\citet{planck2016-l11A} performed a similar thermal dust emission fit
for LFI 70 and 44\,GHz, and found good agreement with the SED at
70\,GHz.  Their 44\,GHz result fits shows a slight excess, while in
this work we find that the template amplitude fit for \BP\ 44\,GHz
(and \WMAP\ Q-band) is also consistent with the thermal dust SED alone, at
least for a synchrotron prior of $\beta_{\mathrm{s}}=-3.1\pm0.1$. Also
for the lower frequency bands we find results consistent with
nonpolarized AME, but with large uncertainties. 

To constrain the corresponding AME polarization fraction, we map out
the corresponding posterior distribution defined by the
change-of-variables in Eq.~\eqref{eq:pfrac_def2}. We create a grid of
potential polarized AME SEDs for different polarization fractions,
$s_{\nu}(p_{\rm AME})$, and calculate a log-likelihood given by
\begin{equation}
-2\ln\mathcal{L}(p_{\rm AME}) = \sum_{\nu} \left(\frac{(\hat{a}_{\nu}-s_{\nu}(p_{\rm AME}))^2}{\sigma_{a,\nu}^2}\right)^2.
\end{equation}
Uncertainties in the thermal dust SED are taken into account by
averaging over random draws from the MBB parameters as described in
Sect.~\ref{sec:sky_model}. Slices of the full posterior distribution for each of the $\beta_{\rm s}$ priors are shown in Fig.~\ref{fig:pmax_ci}, showing the 95\,\% confidence interval for $p_{\mathrm{AME}}^{\rm max}$, as well as the posterior peak. The nominal case (shown in the form of SED constraints in
Fig.~\ref{fig:ame_seds}) is shown in Fig.~\ref{fig:posteriors} in orange. In this particular case, we find that the
posterior peaks below 0\,\% with an upper limit of
$p_{\mathrm{AME}}^{\rm max} < 1.0\,\%$ at the $95\,\%$ confidence
level.

However, as shown in Sect.~\ref{sec:synch}, the current data set has
very little constraining power with respect to
$\beta_{\mathrm{s}}$. We therefore repeat the above calculations for a
grid of $\beta_{\mathrm{s}}$, allowing the prior mean to range between
$\beta=-3.3$ and $\beta=-3.0$. The results from these analyses are
summarized in Fig.~\ref{fig:pmax_ci}. Here we see that for any
$\beta_{\mathrm{s}}\lesssim-3.1$, the distributions are consistent
with a vanishing AME polarization fraction, while for flatter indices
a nominal positive detection emerges. The most likely explanation for
this is that synchrotron emission is oversubtracted by the flat
spectral index, and this may be countered by adding a spurious
polarized AME component. 

Figure~\ref{fig:pmax_ci} summarizes this in the
form of the 95\,\% confidence region as a function of the
$\beta_{\mathrm{s}}$ prior mean. Here we find that the typical upper
limit is $p_{\mathrm{AME}}^{\rm max}\lesssim1.0\,\%$ at 95\,\%
confidence for $\beta_{\mathrm{s}}\lesssim-3.1$, while for
$\beta_{\mathrm{s}}=-3.0$, there is a nominal detection of polarized
AME with $p_{\mathrm{AME}}=2.5\pm1.0\,\%$ at 95\,\% confidence. 

\subsection{Impact of CMB fluctuations}
\label{sub:cmb_test}

To determine the validity of our assumption to ignore the CMB component in Sect.~\ref{sub:sky_model}, we run our analysis on the same set of data, but with an additional CMB component added to each frequency map. This allows us to estimate the affect the CMB component has on our $p_{\rm AME}^{\rm max}$ constraints. 

The CMB component is generated using \texttt{healpy}'s \citep{Zonca2019} \texttt{synfast} routine, using the \Planck\ best-fit $\Lambda$CDM power spectrum\footnote{\url{https://pla.esac.esa.int}} \citep{planck2016-l05}. The results are shown in Fig.~\ref{fig:posteriors}, where we show the $p_{\rm AME}^{\rm max}$ posterior slices of the fiducial analysis in orange and the fiducial-plus-CMB analysis in cyan. We see that an additional set of CMB fluctuations increases the upper limits on $p_{\rm AME}^{\rm max}$ by 10--$15\,\%$. As a result, the main limits quoted in paper are likely overestimated, and therefore conservative, as the addition of unmodeled CMB fluctuations lead to slightly weaker constraints.

\section{Summary}
\label{sec:summary}

We have presented the first constraints on large-scale polarized AME
derived based on the combination of \Planck\ LFI and
\WMAP\ polarization measurements. These constraints are derived within
the context of the \BP\ framework, which allows for detailed
propagation of systematic errors; the current work is an explicit
example of how the \BP\ Monte Carlo frequency map ensembles may be
used for error propagation in external analyses. However, we do note
that only \Planck\ LFI data are currently processed through this
machinery, while traditionally processed maps are employed for
\WMAP. The effect of this mismatch is seen in various residual and
difference maps, clearly demonstrating the presence of instrumental
systematic effects in one or both experiments. Work is currently
on-going to reprocess the \WMAP\ data jointly with LFI \citep{bp17},
but for now we note that the two experiments agree very well in terms
of Galactic estimates, even if their instrumental residuals
differ. We also note that the $\chi^2$ of our fits is
excellent, which indicates that systematic errors are small compared
to the noise level of each experiment. 

We find that the LFI and \WMAP\ data only have limited
constraining power with respect to the spectral index of polarized
synchrotron emission, and any constraint on the AME polarization
fraction is significantly affected by the choice of synchrotron
spectral index prior: The current data set is simply not able to
robustly and simultaneously constrain polarized AME and synchrotron
emission. As a result, we choose to report a limit on the AME
polarization fraction that is explicitly
$\beta_{\mathrm{s}}$-dependent. Specifically, for
$\beta_{\mathrm{s}}\lesssim -3.1$, we find an upper limit of
$p_{\mathrm{AME}}^{\rm max}\lesssim0.6\,\%$ at 95\,\% confidence,
while for $\beta_{\mathrm{s}}\gtrsim -3.1$ there is formally a positive
detection of polarized AME power. 

The maximum polarization fraction of $3.5~\%$ (for $\beta_{\mathrm{s}} = -3.0$) derived here is in good agreement with and more stringent than the large-scale \cite{Macellari_2011} limit of $p_{\mathrm{AME}}<5\%$. Our upper limit also agrees with more stringent upper limits placed in individual molecular cloud complexes, such as those from \cite{battistelli2006,QUIJOTE_II_2016,QUIJOTE_III}, though we emphasize that the AME may be significantly less polarized than the theoretical maximum in these regions based on the magnetic field geometry and level of coherence within the beam and along the line of sight. The nondetection of polarization is in agreement with spinning dust emission theory, driven either by PAHs or nanosilicate grains, but we do not have the sensitivity to discriminate between models with modest levels of polarization and models which predict no polarization. AME driven by emission mechanisms which have polarization fractions of greater than a few percent, such as spinning iron grains \citep{Hoang_2016}, are excluded.

We do note that both \Planck\ 2018 \citep{planck2016-l05} and
\BP\ \citep{bp01,bp14} prefer a steep spectral index of $\boldsymbol{\beta_{\mathrm{s}}} \approx
-3.3$ at high Galactic latitudes. However, those analyses consider only LFI data, and are therefore
susceptible to even stronger degeneracies than what is observed in the
current analysis, which also includes \WMAP\ data. It is conceivable
that the steep spectral index they observe is partially caused by
polarized AME. Another complication is the possibility of curvature in
the synchrotron SED, which has not been addressed in the current
analysis.

The only way to actually break these degeneracies is through
additional high signal-to-noise observations, both by ground-based
low-frequency experiments such as C-BASS \citep{jew2019}, S-PASS \citep{Carretti:2019}, QUIJOTE \citep{genovasantos2015}, and by
next-generation large-scale experiments such as Simons Observatory \citep{SO2019}
and CMB-S4 \citep{cmbs4}. Indeed, we would argue that high-sensitivity
characterization of the low-frequency polarized foreground should be a
primary objective for near-term CMB experiments, as the complexity of
this frequency range can have dramatic consequences for
next-generation CMB B-mode satellite experiments.

This is illustrated in Fig.~\ref{fig:nuref}, which provides an
overview of the polarized microwave sky from 1 to 500\,GHz. Green and
red bands show synchrotron and thermal dust emission as constrained by
\BP\ \citep{bp14}, while the orange curve shows the conservative upper limit on
polarized AME for a synchrotron spectral index prior of $\boldsymbol{\beta_{\mathrm{s}}}=-3.1\pm0.1$, namely
$p_{\mathrm{AME}}^{\rm max}\lesssim1.0\,\%$. Clearly, the presence of
a polarized AME component with an amplitude this level
would be critically important for any future B-mode experiments. In
sum, the current analysis serves as a useful reminder about how little
is still known about low-frequency polarized foregrounds, even after
\Planck\ and \WMAP, and more data are desperately needed.

\begin{acknowledgements}
  We thank Prof.\ Pedro Ferreira and Dr.\ Charles Lawrence for useful suggestions, comments and 
  discussions. We also thank the entire \Planck\ and \WMAP\ teams for
  invaluable support and discussions, and for their dedicated efforts
  through several decades without which this work would not be
  possible. The current work has received funding from the European
  Union’s Horizon 2020 research and innovation programme under grant
  agreement numbers 776282 (COMPET-4; \BP), 772253 (ERC;
  \textsc{bits2cosmology}), and 819478 (ERC; \textsc{Cosmoglobe}). In
  addition, the collaboration acknowledges support from ESA; ASI and
  INAF (Italy); NASA and DoE (USA); Tekes, Academy of Finland (grant
   no.\ 295113), CSC, and Magnus Ehrnrooth foundation (Finland); RCN
  (Norway; grant nos.\ 263011, 274990); and PRACE (EU).
\end{acknowledgements}

\bibliographystyle{aa}

\bibliography{Planck_bib,BP_bibliography,ame}

\end{document}